\documentclass[twocolumn,English,aps,pra,10pt,superscriptaddress,floatfix]{revtex4-2}
\usepackage[colorlinks=true,citecolor=blue,linkcolor=magenta]{hyperref}
\usepackage[normalem]{ulem} 
\usepackage{times}
\usepackage{xcolor}
\usepackage{graphicx}
\usepackage{amssymb}
\usepackage{bm} 
\usepackage{amsfonts,amsmath,amsbsy,latexsym}
\usepackage{gensymb} 
\usepackage{setspace}
\usepackage{float}
\usepackage[markup=blue, authormarkupposition=left]{changes}
\usepackage[english]{babel}
\usepackage{url}
\usepackage{xspace} 
\usepackage{siunitx}
\DeclareSIUnit \dBc {dBc}
\usepackage{soul}
\usepackage{upgreek} 

\DeclareUnicodeCharacter{2082}{$_2$} 
\DeclareUnicodeCharacter{2083}{$_3$} 

\newcommand{\fref}[1]{Fig. \ref{#1}}

\newcommand{\IDT}[0]{interdigital transducer} 
\newcommand{\LT}[0]{$\mathrm{LiTaO}_3$}

\begin{document}

\title{ {High-Efficiency Acousto-Optic Modulation on Non-Suspended Thin-Film Lithium Tantalate}}

\author{Weiran Zhou}
\affiliation{School of Microelectronics, University of Science and Technology of China, Hefei, China}
\affiliation{State Key Laboratory of Materials for Integrated Circuits, Shanghai Institute of Microsystem and Information Technology, Chinese Academy of Sciences, 865 Changning Road, Shanghai 200050, China.}

\author{Chengli Wang}
\email[]{wangcl@mail.sim.ac.cn}
\affiliation{State Key Laboratory of Materials for Integrated Circuits, Shanghai Institute of Microsystem and Information Technology, Chinese Academy of Sciences, 865 Changning Road, Shanghai 200050, China.}
\affiliation{Center of Materials Science and Optoelectronics Engineering, University of Chinese Academy of Sciences,Beijing 100049, China}

\author{Xuqiang Wang}
\affiliation{State Key Laboratory of Materials for Integrated Circuits, Shanghai Institute of Microsystem and Information Technology, Chinese Academy of Sciences, 865 Changning Road, Shanghai 200050, China.}
\affiliation{Center of Materials Science and Optoelectronics Engineering, University of Chinese Academy of Sciences,Beijing 100049, China}

\author{Bowen Chen}
\affiliation{State Key Laboratory of Materials for Integrated Circuits, Shanghai Institute of Microsystem and Information Technology, Chinese Academy of Sciences, 865 Changning Road, Shanghai 200050, China.}
\affiliation{Center of Materials Science and Optoelectronics Engineering, University of Chinese Academy of Sciences,Beijing 100049, China}

\author{Jiachen Cai}
\affiliation{State Key Laboratory of Materials for Integrated Circuits, Shanghai Institute of Microsystem and Information Technology, Chinese Academy of Sciences, 865 Changning Road, Shanghai 200050, China.}
\affiliation{Center of Materials Science and Optoelectronics Engineering, University of Chinese Academy of Sciences,Beijing 100049, China}

\author{Tianyao Yang}
\affiliation{State Key Laboratory of Materials for Integrated Circuits, Shanghai Institute of Microsystem and Information Technology, Chinese Academy of Sciences, 865 Changning Road, Shanghai 200050, China.}
\affiliation{Center of Materials Science and Optoelectronics Engineering, University of Chinese Academy of Sciences,Beijing 100049, China}

\author{Dongchen Sui}
\affiliation{State Key Laboratory of Materials for Integrated Circuits, Shanghai Institute of Microsystem and Information Technology, Chinese Academy of Sciences, 865 Changning Road, Shanghai 200050, China.}
\affiliation{Center of Materials Science and Optoelectronics Engineering, University of Chinese Academy of Sciences,Beijing 100049, China}

\author{Xinjian Ke}
\affiliation{Shanghai Novel Si Integration Technology, Shanghai 201815, China}

\author{Yang Chen}
\affiliation{Shanghai Novel Si Integration Technology, Shanghai 201815, China}

\author{Xudong Wang}
\affiliation{State Key Laboratory of Materials for Integrated Circuits, Shanghai Institute of Microsystem and Information Technology, Chinese Academy of Sciences, 865 Changning Road, Shanghai 200050, China.}
\affiliation{Center of Materials Science and Optoelectronics Engineering, University of Chinese Academy of Sciences,Beijing 100049, China}

\author{Ailun Yi}
\affiliation{State Key Laboratory of Materials for Integrated Circuits, Shanghai Institute of Microsystem and Information Technology, Chinese Academy of Sciences, 865 Changning Road, Shanghai 200050, China.}
\affiliation{Center of Materials Science and Optoelectronics Engineering, University of Chinese Academy of Sciences,Beijing 100049, China}

\author{Shibin Zhang}
\affiliation{State Key Laboratory of Materials for Integrated Circuits, Shanghai Institute of Microsystem and Information Technology, Chinese Academy of Sciences, 865 Changning Road, Shanghai 200050, China.}
\affiliation{Center of Materials Science and Optoelectronics Engineering, University of Chinese Academy of Sciences,Beijing 100049, China}

\author{Chengjie Zuo}
\email[]{czuo@ustc.edu.cn}
\affiliation{School of Microelectronics, University of Science and Technology of China, Hefei, China}

\author{Xin Ou}
\email[]{ouxin@mail.sim.ac.cn}
\affiliation{School of Microelectronics, University of Science and Technology of China, Hefei, China}
\affiliation{State Key Laboratory of Materials for Integrated Circuits, Shanghai Institute of Microsystem and Information Technology, Chinese Academy of Sciences, 865 Changning Road, Shanghai 200050, China.}
\affiliation{Center of Materials Science and Optoelectronics Engineering, University of Chinese Academy of Sciences,Beijing 100049, China}
\affiliation{Shanghai Novel Si Integration Technology, Shanghai 201815, China}

\maketitle

\noindent
{\large\textbf{Abstract}}
\\
\textbf{
Acousto-optic (AO) interactions provide a powerful interface between the microwave and optical domains, enabling functionalities such as optical switching, non-reciprocal propagation and efficient microwave-to-optical transduction. Integrated demonstrations to date have largely relied on thin-film lithium niobate (TFLN), which offers strong piezoelectric response and low optical loss performance. Here, we establish lithium tantalate on insulator (LTOI) as a scalable platform for integrated acousto-optics. LTOI combines intrinsically low birefringence, high optical damage threshold, strong electro-optic and Kerr nonlinearities, and superior acoustic quality factors with a mature high-volume manufacturing base. We demonstrate for the first time acousto-optic modulation on {the} LTOI platform. By exploiting the anisotropy of surface acoustic waves, we reveal a direct correlation between acousto-optic modulation efficiency and the electromechanical coupling coefficient of lithium tantalate. In particular, acoustic excitation along the crystal Z-axis enhances the higher-order R1 mode, yielding the highest modulation efficiency.
Our Mach–Zehnder interferometers achieve a modulation efficiency of 0.68 $\mathrm{\mathbf{V \cdot cm}}$, while racetrack resonators reach 0.022 $\mathrm{\mathbf{V \cdot cm}}$—representing, to the best of our knowledge, the lowest $\mathrm{V_\pi L}$ demonstrated in non-suspended ferroelectric platforms. This record performance directly enables microwave-to-optical conversion without suspended structures, establishing LTOI as a robust and scalable platform for integrated acousto–optics with broad applications in communications, signal processing, and quantum information technologies.
}

~\\
\\
\noindent
{\large\textbf{Introduction}}

Acousto-optic (AO) interactions provide a versatile interface between microwave and optical domains, enabling functionalities that are otherwise challenging to achieve in integrated photonics. In particular, they allow for optical switching with high extinction ratios, as well as unique opportunities for non-reciprocal light propagation, broadband spectral filtering, and highly efficient microwave-to-optical transduction \cite{tian2020hybrid,schrodel2024acousto,harris1969acousto,jiang2020efficient,chen2025intermodal}. Such capabilities are of immediate relevance to advanced optical communication, microwave photonic signal processing, and quantum information technologies \cite{lin2025optical,mirhosseini2020superconducting,forsch2020microwave}. To translate these opportunities into complex and large-scale photonic circuits, it is crucial to realize efficient AO modulation on scalable and stable integrated platforms, which enable dense integration of multiple devices with high reproducibility beyond what discrete, or bulk or suspended structures can offer. Thin-film piezoelectric materials such as AlN, GaAs, and TFLN have been widely explored for integrated AOs, where strong mechanical wave excitation and effective coupling with nanophotonic circuits were demonstrated \cite{vainsencher2016bi,bochmann2013nanomechanical,fan2013aluminum,forsch2020microwave,balram2016coherent,jiang2019lithium,cai2019acousto,xu2025silicon}. Among these, TFLN is regarded as a promising platform for AO interaction owing to its large electro-optic coefficient, wide optical transparency window, low optical loss, and strong piezoelectric response enabling efficient acoustic excitation \cite{andrushchak2009complete,weis1985lithium}. Nevertheless, practical implementations can be influenced by photorefractive effects and thermal drift, which affect device stability especially under high optical intensities \cite{chen1969optically,xu2021mitigating,holzgrafe2024relaxation}.

Lithium tantalate (\LT. LT), a ferroelectric crystal structurally similar to LN, has recently emerged as a compelling alternative \cite{wang2024lithium,yu2024tunable,he2025lithium,zhang2025ultrabroadband,wang2024ultrabroadband}. LT exhibits equal, and in some cases superior, properties compared to LN—including intrinsically lower birefringence, higher optical damage threshold, and the coexistence of strong electro-optic and Kerr nonlinearities \cite{wang2024lithium,jacob2004temperature,ccabuk1999urbach}. In addition to these photonic merits, LT~also offers superior acoustic and mechanical properties, such as high quality factors and excellent temperature stability in surface acoustic wave (SAW) devices \cite{zhang2023high,wu2022advanced,naumenko2018multilayered,naumenko2021temperature}.  {These attributes establish LT as the dominant material for commercial radio frequency (RF) SAW filters, especially in 5G/6G consumer electronics. The thin-film LT platform (industrially referred to as Piezoelectric-on-Insulator or POI) has already achieved high-volume manufacturing maturity. Major substrate manufacturers have established mass production lines supplying millions of wafers annually to RF component leaders like Murata and Qualcomm \cite{soitec2024urd}. This existing industrial ecosystem, with annual thin-film wafer production exceeding 750,000 units, provides LTOI with a distinct scalability advantage over other ferroelectric platforms.} This unique combination of excellent optical and acoustic properties, together with a mature and high-volume industrial manufacturing base, makes LT {a potentially} outstanding platform for exploring AO interactions with broad technological relevance. 

While LTOI is extensively studied for high-performance acoustic filters, more recently for a variety of integrated photonic devices\cite{wang2024lithium,wang2024ultrabroadband,zhang2025ultrabroadband}, investigations into AO interactions on this platform so far remain unexplored. Here, we demonstrate for the first time AO modulation on LTOI. By systematically engineering the in-plane orientation of SAW devices, we exploit the anisotropy of acoustic modes to achieve enhanced acoustic quality factors and efficient AO modulation. Importantly, we realize these results in non-suspended LTOI structures, providing mechanically stable yet efficient AO interaction. Our devices exhibit modulation efficiencies of 0.68 $\mathrm{V\cdot cm}$ in Mach–Zehnder interferometers and 0.022 $\mathrm{V\cdot cm}$ in racetrack resonators, the latter further enabling direct microwave-to-optical conversion without requiring suspended structures. These results establish LTOI as a practical and scalable platform for high-performance AOs, paving the way toward stable, broadband, and multifunctional microwave photonic circuits and systems.

\begin{figure*}[htp] 
	\centering
	\includegraphics[scale = 1]{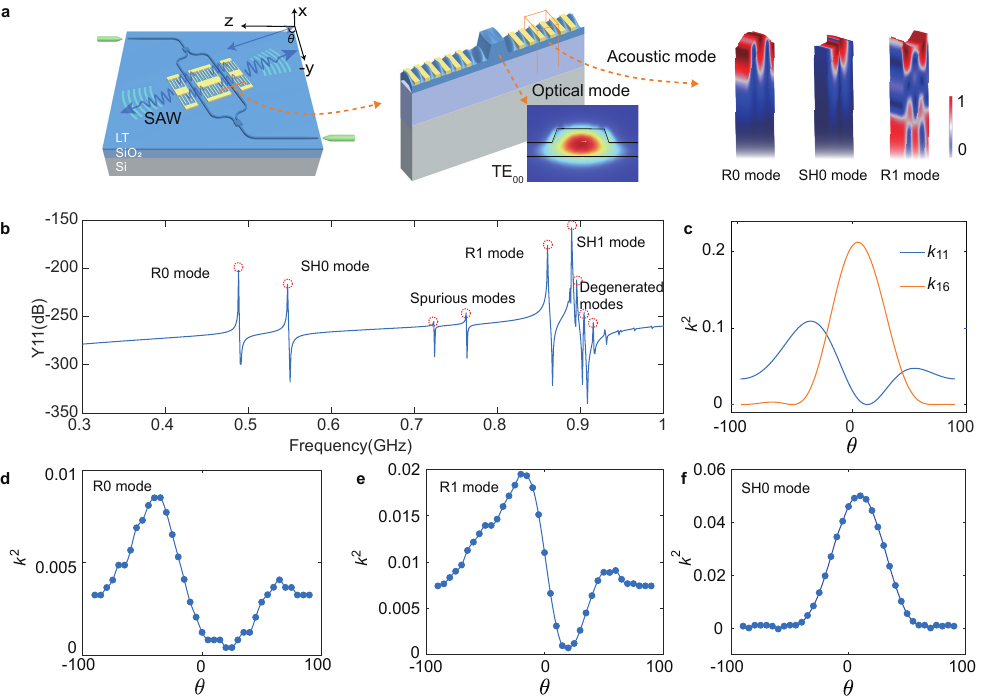}
	\caption{\textbf{Design and simulation.}
  (a) Schematic layouts of AO modulators. Light is coupled to/from the chip by inverse taper (green). An SAW is launched by the IDT (yellow), and the amplitude is enhanced in cavity formed by the reflectors. Cross-sectional renderings of the optical and acoustic fields are also shown. {The spacing between the MZI arms is designed to match the anti-symmetric acoustic phase condition for push-pull modulation.} (b) Obtained Y11 curve from FEM simulation of acoustic resonator on LTOI.  {The Y11 magnitude is normalized to 1 Siemens (0 dB = 1 S).} (c) Static $k^2$ of X-cut LT, $k_{11}$ represents $k^2$ of Rayleigh wave, $k_{16}$ represents $k^2$ of shear horizontal wave. (d) Effective $k^2$ of Rayleigh type SAW R0 varies with propagation angle. (e) Effective $k^2$ of Rayleigh type SAW R1 varies with propagation angle. (f) Effective $k^2$ of Rayleigh type SAW SH0 varies with propagation angle.
}
	\label{fig1}
\end{figure*}

{\large\textbf{Device design and fabrication}}

The schematic of a Mach-Zehnder interferometer (MZI)-type AO device is shown in \fref{fig1}(a). {All the devices in this work were fabricated on a lithium tantalate on insulator (LTOI) wafer, consisting of a 600 nm X-cut LT thin film, a 4.7 $\mu m$ buried oxide (BOX) layer, and a 525 $\mu m$ silicon substrate. The LTOI wafer was fabricated using the ion-slicing and wafer bonding technique. Commercially available X-cut LT bulk crystals were implanted with He+ ions and bonded to a silicon handle wafer with thermal oxide, followed by exfoliation and chemical mechanical polishing (CMP) to achieve the target thickness.} Light is coupled into the waveguide by an inverse taper, while the SAWs are excited by applying an RF signal to the \IDT~through a GSG probe. In this design, the SAWs propagate perpendicular to the two arms of the MZI and are well confined and resonated by grating electrode to form an acoustic cavity. {The modulator operates based on a push-pull mechanism. Since the acoustic velocity ($v_{a} \approx 4000 m/s \sim 8000 m/s$) is orders of magnitude slower than the optical group velocity ($v_{g}$), the acoustic wave acts as a quasi-static periodic strain field during the optical transit time. To maximize modulation efficiency, we employed spatial phase engineering by setting the distance between the two MZI arms to correspond to an odd multiple of half the acoustic wavelength ($D \approx (n + 1/2)\Lambda_{acoustic}$). Consequently, the acoustic phases at the two arms differ by $\pi$, inducing refractive index changes of opposite signs simultaneously (differential modulation) and doubling the total phase shift.} When interacting with the waveguide, the guided optical field—typically the fundamental TE mode—can be modulated through several mechanisms: the electro-optic effect, the photoelastic effect, and the moving-boundary effect \cite{mytsyk2021photoelastic}. Unlike the optical waveguide, which usually supports a single dominant mode, the acoustic domain exhibits multiple mechanical modes, such as Rayleigh (R) and shear-horizontal (SH) modes. {As illustrated on the right side of Fig. 1, we employed a unit cell finite-element model (FEM) incorporating an interdigital transducer and periodic boundary conditions (PBC). This approach allows for the efficient extraction of eigenfrequencies, mode profiles, and electromechanical coupling coefficients without the prohibitive computational cost of simulating the entire macroscopic cavity. Using this model, we find various acoustic modes in this LTOI platform}, which can be identified as R0, SH0 and R1 mode according to their displacement mode profiles \cite{wu2024comparative,peng2024strain,huang2018influence}. The corresponding resonance frequencies of these SAW modes are displayed in the simulated admittance curve in \fref{fig1}(b), where multiple acoustic modes and their higher-order harmonics are clearly identified and labeled. Additional small resonances are also observed, which can be attributed to spurious modes. These originate from non-ideal lateral electric-field distributions in the IDT, which couple to unintended acoustic polarizations or transverse standing-wave modes. Such spurious responses are well known in acoustic filter engineering and are often mitigated by optimizing IDT geometry to suppress their excitation or to avoid degeneracy with the primary modes \cite{stettler2023transversal,kadota2018suprious}. Importantly, all of these acoustic modes are capable of coupling to the guided optical mode, leading to diverse modulation effects across different frequency bands.

\begin{figure*}[ht]
	\centering
	\includegraphics[scale = 1]{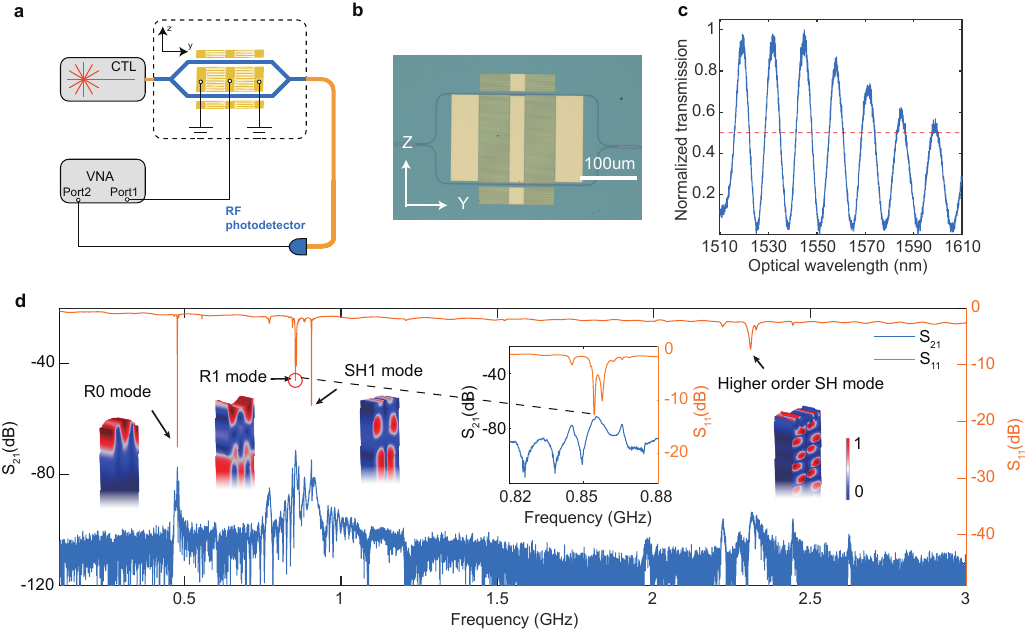}
	\caption{\textbf{Characterization of the AO MZI.}
  (a) Simplified experimental schematic. CTL, continuously tunable laser; PD, photodetector; The sensitivity of the PD is 22.5 $\mathrm{V/W}$; VNA, vector network analyzer. (b) Microscope image of the fabricated MZI AO device. (c) Optical transmission of the AO MZI. (d) Measured acoustic $\mathrm{S_{11}}$ and opto-acoustic $\mathrm{S_{21}}$ spectra.
}
	\label{fig2}
\end{figure*}

Having identified the dominant acoustic modes in the LTOI platform, we next analyze their electromechanical coupling coefficients ($k^2$), which quantify the efficiency of converting electrical energy into acoustic energy and are thus critical for achieving strong AO modulation. The $k^2$ is defined as the ratio of interactional to mechanical and dielectric energy densities, and it directly reflects the strength of piezoelectric transduction \cite{hashimoto2000surface}:
\begin{center}
 $k^2~=~\displaystyle{\frac{U_I}{U_MU_E}}$
\end{center}
where $U_I$, $U_M$, $U_E$~refer to interactional, mechanical and dielectric energy density. Notably, the $k^2$ values of the aforementioned acoustic modes exhibit strong anisotropy, particularly in X-cut LT, a property that has been extensively studied in the context of acoustic resonators \cite{naumenko2019high, assila2019high,patel2010investigation}.
Such anisotropic behavior directly impacts the achievable AO modulation efficiency \cite{shao2022electrical,sarabalis2020acousto,shi2025observation}. To provide a simplified illustration of the anisotropic geometry, we define the –Y axis as the reference propagation direction ($\theta = 0^{\circ}$) for SAW excitation. The static $k^2$ values of the Rayleigh and shear horizontal modes, denoted as $k_{11}$ and $k_{16}$, respectively, can be calculated from the piezoelectric, dielectric, and elastic constants of the material \cite{chang2002analysis} (calculation details provided in SI, Part 1). As shown in \fref{fig1}(c), the $k^2$ of the Rayleigh mode reaches a maximum of approximately 11\% near $\theta = -30^{\circ}$, whereas the $k^2$ of the SH0 mode peaks at about 21\% near $\theta = 10^{\circ}$.  {These values are comparable to those of lithium niobate (LN), confirming the strong piezoelectric nature of the material. Furthermore, they are over an order of magnitude higher than other common integrated photonic materials such as aluminum nitride (AlN, $k^2 \approx 0.3-1\%$) and gallium arsenide (GaAs, $k^2 \approx 0.07\%$). While LN exhibits slightly higher maximum coupling coefficients in specific orientations, LT provides a superior balance of electromechanical coupling, acoustic quality factor (Q), and thermal stability (TCF) \cite{takai2017ihp,tanaka2024evolution}, as detailed in Supplementary Note S1. These large $k^2$ values indicate the strong potential of LT for efficient piezoelectric transduction...} While the static analysis provides valuable insight into intrinsic anisotropy, it does not account for the effects of multilayer device geometry. We therefore employ full 3D FEM simulations on the LTOI stack to further evaluate the $k^2$ of key acoustic modes, including the fundamental Rayleigh (R0), first-order Rayleigh (R1), and fundamental shear-horizontal (SH0) modes. The results, shown in \fref{fig1}(d,e,f) reveal angular dependencies similar to the static case, but with reduced absolute values. This reduction originates from acoustic energy leakage into the buried silicon dioxide buffer layer, which has a relatively low acoustic velocity \cite{zhang2023high}. Such losses, however, can be mitigated by employing alternative buffer materials with higher acoustic velocities, such as silicon carbide or sapphire, as recently demonstrated in advanced acoustic resonator designs \cite{li2022high, wu2022advanced}.

To experimentally investigate the diverse AO interactions and their anisotropic behavior on LTOI platform, MZI-type AO devices were fabricated. The waveguide patterns were defined by electron-beam lithography (EBL) and etched to a depth of 300 nm using $\mathrm{Ar^+}$ ion-beam etching (IBE) at an incidence angle of $70^{\circ}$. The waveguide width was set to 1.2~$\upmu$m. The operating frequency of the SAWs was chosen below 1 GHz to match the acoustic mode dimensions with those of the fundamental optical mode and to suppress acoustic energy leakage into the buffer layer at higher frequencies. The IDTs were designed with a wavelength of 6~$\upmu$m and an electrode width of 1.5~$\upmu$m. {The IDT comprises 30 finger pairs, a number chosen to optimize impedance matching to the 50 $\Omega$ source. Furthermore, acoustic grating reflectors are employed on both sides of the IDT structure. These reflectors are designed to form a Fabry-Perot cavity that confines the acoustic waves, thereby significantly enhancing the acoustic energy density in the vicinity of the optical waveguide.} The electrode patterns were defined in AZ5214 photoresist and developed to form lift-off templates. A Cr/Au bilayer (10 nm/80 nm) was deposited by electron-beam evaporation, followed by acetone lift-off to obtain low-resistance electrodes. Finally, the chips were diced and assembled into the coupling structure.
~\\
\\
{\large\textbf{MZI-type AO modulator and its anisotropy}}
\begin{figure*}[ht]
	\centering
	\includegraphics[scale = 0.9]{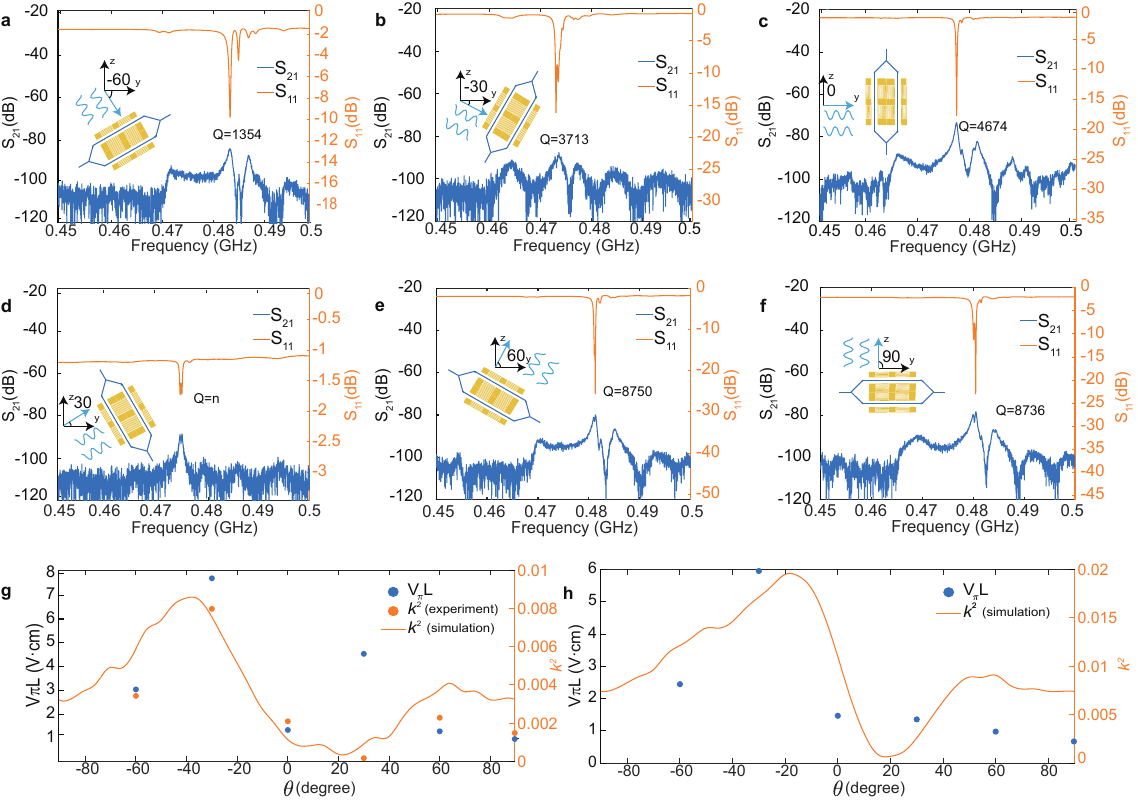}
	\caption{\textbf{Characterization of the anisotropy of AO modulation in Lithium Tantalate.}
 (a) Measured $\mathrm{S_{11}}$ and $\mathrm{S_{21}}$ spectrum of R0 mode when $\theta$~=~-60$\degree$. (b) Measured $\mathrm{S_{11}}$ and $\mathrm{S_{21}}$ spectrum of R0 mode when $\theta$~=~-30$\degree$. (c) Measured $\mathrm{S_{11}}$ and $\mathrm{S_{21}}$ spectrum of R0 mode when $\theta$~=~0$\degree$. (d) Measured $\mathrm{S_{11}}$ and $\mathrm{S_{21}}$ spectrum when $\theta$~=~30$\degree$. (e) Measured $\mathrm{S_{11}}$ and $\mathrm{S_{21}}$ spectrum when $\theta$~=~60$\degree$. (f) Measured $\mathrm{S_{11}}$ and $\mathrm{S_{21}}$ spectrum when $\theta$~=~90$\degree$. (g) Measured $\mathrm{V_\pi L}$ and $k^2$ varies with $\theta$ of R0 mode. (h) Measured $\mathrm{V_\pi L}$ and $k^2$ varies with $\theta$ of R1 mode.
}
	\label{fig3}
\end{figure*}

Prior studies have explored several strategies to enhance AO coupling efficiency, such as employing suspended structures to improve acoustic confinement \cite{shao2019microwave}, using high–photoelastic materials (e.g. $\mathrm{As_2S_3}$) to enhance the photoelastic effect \cite{wan2022highly}, or using optical microcavities to boost photon–phonon interaction \cite{wan2025hybrid}. {However, relatively little attention has been paid to exploiting the crystalline anisotropy and device geometry to optimize the excitation and selection of acoustic modes, even though coupling efficiency can vary significantly between modes\cite{shi2025observation}.} In particular, higher-order acoustic modes operate at higher frequencies than fundamental modes, underscoring the importance of acoustic mode engineering for pushing AO devices toward higher-frequency operation. To investigate the role of acoustic modes in AO modulation, we characterized the MZI-type modulator using the setup illustrated in \fref{fig2}(a).
 {In our devices, the spacing between the waveguide and the electrodes was experimentally optimized to maximize the acousto-optic overlap integral (see Supplementary Information Part 4), thereby enhancing the AO modulation efficiency specifically for the targeted modes.} Light was coupled into and out of the devices via lensed fibers aligned to inverse tapers, yielding a fiber-to-fiber insertion loss of 12 dB (6 dB per facet). The optical transmission response of the unbalanced MZI with 15 nm period agrees well with the expected optical path difference, as shown in \fref{fig2}(b). To experimentally optimize the modulation efficiency, the laser wavelength was set to 1530 nm, corresponding to 50$\%$ transmission point and a $\frac{\pi}{2}$ phase difference between the two arms.

\begin{figure*}[ht]
	\centering
	\includegraphics[scale = 1]{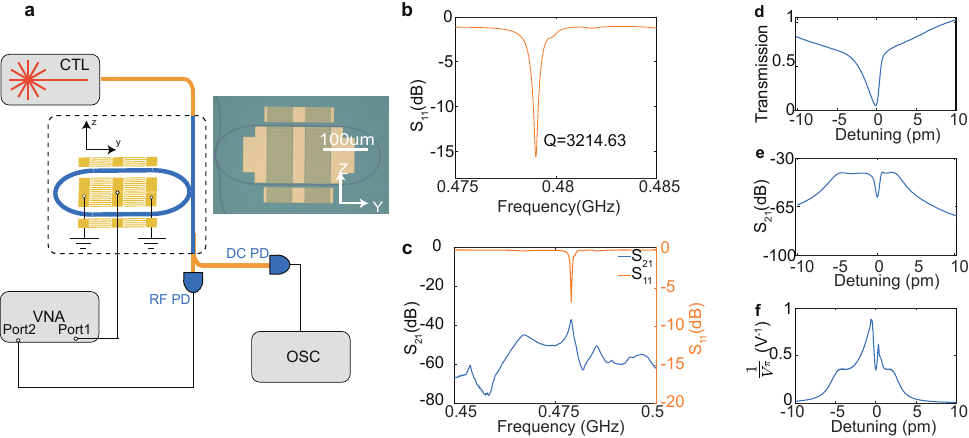}
	\caption{\textbf{Characterization of the AO racetrack resonator.}
  (a) Simplified experimental schematic. CTL, continuously tunable laser; PD, photodetector; VNA, vector network analyzer; OSC, oscilloscope. (b) $\mathrm{S_{11}}$~reflection spectrum of the acoustic resonator. (c) Measured acoustic $\mathrm{S_{11}}$~and opto-acoustic $\mathrm{S_{21}}$~spectra from the same setup as that employed for the MZI AO modulators. (d) Optical transmission varies with the detuning of racetrack cavity. (e) $\mathrm{S_{21}}$~varies with the detuning of racetrack cavity. (f) $V_\pi$~varies with the detuning of racetrack cavity.
}
	\label{fig4}
\end{figure*}

The modulation efficiency was evaluated from the measured $\mathrm{S_{21}}$ spectra. As schematically illustrated in \fref{fig2}(a), port 1 of the VNA drives the \IDT~of the acoustic resonator, while Port 2 monitors the signal from the photoreceiver. \fref{fig2}(d) presents the measured $\mathrm{S_{21}}$ and $\mathrm{S_{11}}$ spectra, revealing multiple resonance peaks corresponding to different acoustic modes. As evidenced by the peaks in the $\mathrm{S_{11}}$ spectra, the frequencies of the acoustic modes show excellent agreement with the simulation results. Specifically, distinct acoustic resonances are identified at approximately 0.478 GHz, 0.858 GHz, 0.895 GHz, and 2.38 GHz, corresponding to the R0, R1, SH1, and a higher-order SH mode, respectively. Notably, the SH0 mode was not effectively excited at $\theta$~=~90$\degree$, a phenomenon that was also predicted by FEM simulations as shown in Supplementary Information (SI) Part 2. {Notably, fine resonance splitting is observed surrounding the main peak, particularly at angles where the electromechanical coupling is maximized (e.g., near $-30^\circ$ for the R1 mode). This phenomenon is attributed to the superposition of intrinsic spatially degenerate modes and a specific longitudinal cavity mode induced by the acoustic reflectors. As validated by FEM simulations in Supplementary Note S7, the reflectors introduce a high-Q cavity mode (simulated at 0.865 GHz) situated between the inherent degenerate modes (0.852 GHz and 0.889 GHz). These fine cavity features are spectrally resolved only when the anisotropic electromechanical coupling is sufficiently strong to efficiently excite the standing waves.}

It is observed that each resonance peak in the $\mathrm{S_{11}}$ spectrum corresponds to a clear peak in the $\mathrm{S_{21}}$ response, indicating efficient AO modulation for multiple acoustic mode \fref{fig2}(c). Among them, the R1 mode exhibits the strongest modulation efficiency, attributable {to its maximal overlap with the optical mode. Our quantitative analysis of the acousto-optic overlap integral confirms that the R1 mode exhibits a higher overlap factor compared to the R0 and SH0 modes, as detailed in Supplementary Note S8 and Figure S8.} The half-wave voltage ($\mathrm{V_\pi}$) of the MZI was extracted from the $\mathrm{S_{21}}$ measurements using the derived relationship \cite{shao2019microwave}:
\begin{center}
 $\mathrm{S_{21}}~=~(\displaystyle{\frac{\pi {\mathrm{R_{PD}}} {\mathrm{I_{rec}}}}{V_\pi }})^2$
\end{center}

 {Where $R_{PD}$ and $I_{rec}$ refer to the photoreceiver responsivity and the detected optical power, respectively.} At the R1 resonance (0.855 GHz), we extracted $\mathrm{V_\pi}~=~ 34.17$ V, corresponding to a $\mathrm{V_\pi L}$ product of 0.68~$\mathrm{V \cdot cm}$. Systematic investigation and comparison of different acoustic modes are therefore essential to understand the underlying AO interactions. Moreover, the simultaneous excitation of multiple acoustic modes on the LTOI platform enables multi-band modulation within a single device, suggesting promising opportunities for wavelength-division multiplexing in integrated microwave–photonic circuits.

To investigate the anisotropy of AO modulation, six devices with orientations of –60 $\degree$, –30$\degree$, 0$\degree$, 30$\degree$, 60$\degree$ and 90$\degree$ were fabricated. Due to the presence of degenerate modes in the R1 mode (as predicted by FEM simulations) and the relatively lower modulation efficiency of shear modes compared to Rayleigh modes \cite{li2023frequency}, the fundamental Rayleigh mode (R0) was selected for initial analysis to ensure unambiguous modal characterization. The $\mathrm{S_{11}}$ and $\mathrm{S_{21}}$~spectra were acquired in the frequency range of 0.45–0.50 GHz $[$\fref{fig3}(a-f)$]$, where the R0 mode is prominent. The extracted modulation efficiencies are summarized in \fref{fig3}(g), showing a strong dependence of AO modulation efficiency on propagation direction. The $k^2$ of the R0 mode was extracted from measurements and shows good agreement with FEM simulations $[$\fref{fig3}(g)$]$. The acoustic quality factor $Q_{\mathrm{m}}$, derived from the $\mathrm{S_{11}}$~resonance dips, reached a maximum of 8750, indicating low acoustic loss and strong confinement in the non-suspended LT structure. Such high $Q_{\mathrm{m}}$ factor further confirm efficient AO coupling, underscoring the potential of LT for high-performance integrated AO devices.

Based on the standard extraction method of 
$\mathrm{V_\pi L}$~for MZI AO modulators, the R0 values at orientations of –60 $\degree$, –30$\degree$, 0$\degree$, 30$\degree$, 60$\degree$ and 90$\degree$ were determined to be 3.053, 7.702, 1.357, 4.542, 1.303, and 0.979 $\mathrm{V \cdot cm}$, respectively. The trend of $\mathrm{V_\pi L}$ correlates well with non-zero $k^2$, where lower $\mathrm{V_\pi L}$ values coincide with higher $Q_\mathrm{m}$. This inverse relation between $k^2$ and $Q_\mathrm{m}$ is consistent with the well-established acoustic figure of merit (FOM), defined as $k^2 \cdot Q$ \cite{wu2022advanced,wu2024comparative}. {Among all orientations, the R0 mode was effectively excited except at $30^{\circ}$, where excitation was suppressed. This suppression arises from the intrinsic anisotropy of the X-cut LT crystal. As predicted by our FEM simulations (Fig. 1d) and theoretical calculations, the effective electromechanical coupling coefficient ($k^2$) for the R0 mode passes through a null near this specific propagation angle due to the rotation of the piezoelectric tensor.} Under these conditions, the stored acoustic energy density is primarily governed by $Q_\mathrm{m}$, yielding an inverse proportionality between modulation efficiency and $k^2$. A similar characterization was performed for the R1 mode (see Supplementary Information, part 5), revealing a consistent trend between modulation efficiency (represented by 1/($\mathrm{V_\pi L}$)) and the simulated $k^2$ \fref{fig3}(h). {However, the presence of acoustic reflectors in our design creates a high-Q cavity that supports multiple longitudinal modes spectrally close to the main R1 resonance. This results in significant spectral crowding and overlapping peaks in the measured $S_{11}$ admittance spectrum. These overlapping resonances prevent the unambiguous identification of the distinct series ($f_s$) and parallel ($f_p$) frequencies required for the standard extraction method or accurate fitting using the Modified Butterworth-Van Dyke (mBVD) model. Consequently, a reliable experimental determination of $k^2$ for this mode was not feasible, and we relied on FEM simulations for this specific parameter.} {Systematic investigation and comparison of different acoustic modes are therefore essential to understand the underlying AO interactions. Numerical simulations reveal that the spatial profile of the R1 mode leads to the most significant refractive index change within the waveguide core (see SI for details), directly resulting in the record-low $V_{\pi}L$ product.} The highest modulation efficiency is achieved at a crystal orientation of 90$\degree$. This result provides essential insight for designing highly efficient AO modulators and underscores the applicability of LT in advanced photonic systems.

~\\
\\
{\large\textbf{Racetrack-type AO modulator}}

Compared with MZI structures, racetrack modulators exhibit higher modulation efficiencies owing to the resonant enhancement of photon–phonon coupling in the optical cavity. The small footprints of the acoustic resonators are more suitable for racetrack cavity than for MZI structures. The fabricated racetrack cavity achieved an optical quality factor ($Q_\mathrm{L}$) of $3.17 \times 10^5$ for the TE-polarized mode at a wavelength of 1552.7 nm (see Supplementary Information, Part 6). Initial characterization was performed using the same setup as that employed for the MZI AO modulators. The strongest acoustic response {occurs} at 0.478~GHz, corresponding to a higher-order Rayleigh R0 mode with an acoustic quality factor as high as 3214 $[$\fref{fig4}(b)$]$. Microwave-to-optical transduction was assessed via $\mathrm{S_{21}}$ spectroscopy \fref{fig4}(c). At the 0.478 GHz resonance, $\mathrm{S_{21}}$ = –35.4 dB was measured {at the optimized optical power.} From these data, a half-wave voltage $\mathrm{V_\pi}$ of 1.2 V was extracted. Normalizing by the resonator length of 200 $\upmu$m yields a $\mathrm{V_\pi L}$ product of 0.024 V·cm, confirming the strong modulation efficiency enabled by racetrack resonators.

In racetrack modulators, the modulation-induced change in optical intensity is inherently linked to the resonance characteristics of the cavity transmission spectrum \cite{wan2025hybrid}. The modulation strength therefore depends on cavity detuning, i.e., the laser wavelength relative to the cavity resonance. To systematically probe this effect, we constructed a setup comprising a CTL, a VNA, a PD, and an OSA, allowing synchronous acquisition of optical transmission and $\mathrm{S_{21}}$ \fref{fig4}(a). The input wavelength was swept by the CTL while the VNA was fixed at 0.478 GHz. {As shown in \fref{fig4}(d,e,f), the photon-phonon coupling is maximized under blue detuning of the cavity. This phenomenon is governed by the interplay between cavity optomechanical dynamical back-action and thermo-optic stability. Theoretically, in the blue-detuned regime ($\Delta > 0$), the optomechanical interaction favors Stokes scattering, where the scattering of pump photons into the cavity resonance is accompanied by phonon emission, effectively amplifying the acoustic oscillation \cite{jiang2019lithium}.} Under blue-detuned conditions, a minimum $\mathrm{V_\pi}$ of 1.1 V is obtained, corresponding to a record-low $\mathrm{V_\pi L}$ of 22 mV·cm for the 200-$\upmu$m resonator.

~\\
\\
{\large\textbf{Microwave-to-Optical conversion}}
\begin{figure*}[htb]
	\centering
	\includegraphics[scale = 1]{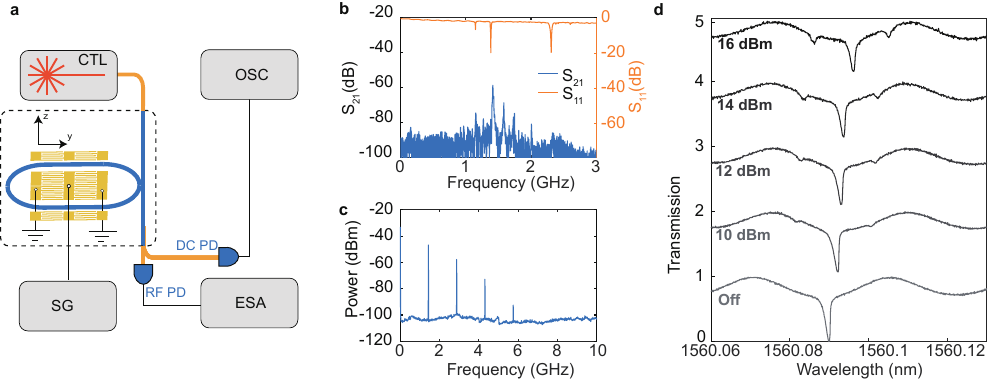}
	\caption{\textbf{Demonstration of a microwave-photonic link.}
  (a) Simplified experimental schematic. CTL, continuously tunable laser; PD, photodetector; SG, signal generator; OSC, oscilloscope; ESA, spectrum analyzer. (b) Measured acoustic $\mathrm{S_{11}}$ and opto-acoustic $\mathrm{S_{21}}$ spectra from the same setup as that employed for the MZI AO modulators. (c) Microwave spectrum of the PD output signal with a microwave power of 16 dBm applied to the \IDT~of acoustic resonator. The laser is blue-detuned. (d) Optical transmission of the racetrack cavity for different microwave powers. The emerging multiple dips represent the coupling of acousto-optic modulation sidebands into the optical cavity modes.
	}
	\label{fig5}
\end{figure*}
\begin{table*}[ht]
\centering
\caption{Comparison of AO modulation Performance of Previously Published Studies}
\label{tab:modulator_comparison}
\begin{tabular}{@{}l l c c S[table-format=1.3] S[table-format=4] S[table-format=1.3e1] S[table-format=1.3] S[table-format=4]@{}}
\toprule
\textbf{Ref.} & \textbf{Platform} & \textbf{Suspended} & \textbf{Type} & \textbf{Frequency[~GHz]} & \textbf{$Q_{\mathrm{m}}$} & \textbf{$Q_{\mathrm{L}}$} & \textbf{$\mathrm{V_\pi L}$[~$\mathrm{V \cdot cm}$]} \\
\cite{huang2025acousto} & AlN on Si & n & FP & 2.250 & $\ /$& $\ /$ & 0.330 \\
\cite{sarabalis2020acousto} & X-cut LN on sapphire & n & MZI & {$0.700-0.800$} & 600 & $\ /$ & 0.940\\
\cite{shao2019microwave} & X-cut LN & y & MZI & 3.330 & 3600 & $\ /$ & 0.046\\
\cite{shao2019microwave} & X-cut LN & y & racetrack &  2.000 & 1562 & {$2.2\times 10^6$} & 0.0077 \\
\cite{xu2024unveiling} & Si on X-cut LNOI & n & MZI &  0.496 & $\ /$ & $\ /$ & 0.496 \\
\cite{khan2020extraction} & $\mathrm{As_2S_3}$/SiO2/Y-cut LN& n & MZI &  0.11 & $\ /$& $\ /$ & 0.94 \\
\cite{wan2022highly} & As2S3 on X-cut LN & n & MZI & 0.84 & 202.81 & $\ /$ & 0.03 \\
\cite{wan2025hybrid}& As2S3 on X-cut LN & n & MZI & 0.84 & $\ /$& {$2.58\times 10^5$} & 0.009 \\
This work & X-cut LTOI & n & MZI & 0.855 & 8750 & $\ /$ & 0.680 \\
This work & X-cut LTOI & n & racetrack & 0.478 & 3214 & {$3.17\times 10^5$} & 0.022\\
This work & X-cut LTOI & n & racetrack & 1.441 & 921 & {$5.2\times 10^5$} & 0.048 \\
\end{tabular}
\end{table*}

The AO modulators developed in this work enable applications in microwave-to-optical conversion \cite{blesin2024bidirectional,yang2025multi}, a function critical for low-loss long-haul microwave signal transport and on-chip photonic signal processing. To demonstrate this capability, a racetrack AO modulator operating at a resonant frequency of 1.438 GHz was designed and fabricated (see Supplementary Information, Part 3). The electrode width was reduced to 500 nm to raise the resonant frequency of the acoustic cavity. {The SAW resonator is formed by an interdigital transducer (IDT) situated between two acoustic reflectors. This configuration creates a Fabry-Perot-like acoustic cavity that confines and enhances the acoustic energy, thereby boosting the acousto-optic modulation efficiency.} The optical racetrack achieved a loaded quality factor $Q_\mathrm{L}$ of $5.2 \times 10^5$ for the TE-polarized mode at 1559 nm (see Supplementary Information, Part 6). The device was characterized using the same setup as for the MZI modulators. The measured $\mathrm{S_{11}}$ and $\mathrm{S_{21}}$ spectra are shown in \fref{fig5}(b). At the 1.441 GHz resonance, $\mathrm{S_{21}}$ reached –58.6 dB. From these measurements, a half-wave voltage $\mathrm{V_\pi}$ of 2.4 V was extracted. Given the device length of 200 um, the resulting $\mathrm{V_\pi L}$ product is calculated to be 0.048 ~$\mathrm{V \cdot cm}$.

To evaluate performance of our device in a microwave photonic link, a system comprising an electrical spectrum analyzer (ESA), a PD, a signal generator (SG) and an OSC is constructed \fref{fig5}(a). {The periodic ripples observed in the $S_{21}$ spectrum (b) correspond to the longitudinal modes of the acoustic cavity.} The device was subsequently integrated into the microwave photonic link. The generated optical sidebands under microwave drive are shown in \fref{fig5}(d), {two resonance dips appear in the optical transmission spectrum as the RF driving power increases from Off to 16 dBm. These dips are identified as modulation sidebands generated at frequencies $f_{opt} \pm n \cdot f_{a}$ (where $f_{opt}$ is the optical carrier frequency and $f_{a}$ is the acoustic frequency). In the high-Q racetrack resonator, these sidebands couple into the cavity resonance modes during the laser wavelength scan. The increasing depth and number of these sidebands with higher RF power directly demonstrate the high microwave-to-optical conversion efficiency of the non-suspended LTOI platform.} A small resonance shift is observed during spectrum scanning, attributable to photorefractive effects, which are known to be more pronounced in LN than in LT \cite{shao2019microwave}. 

~\\
\\
{\large\textbf{Discussion and outlook}}

The performance of AO modulators implemented on the LTOI platform is summarized in Table I, which highlights key metrics such as half-wave voltage–length product ($\mathrm{V_\pi L}$), acoustic quality factor ($Q_{\mathrm{m}}$), and operational frequency. Through systematic investigation of the anisotropic properties of LT, we demonstrate for the first time highly efficient AO modulation on this platform. By integrating optical racetrack resonators, we further achieve efficient microwave-to-optical conversion.

As shown in \ref{tab:modulator_comparison}, our devices exhibit performance that is highly competitive with other AO platforms. Compared with earlier results on LN-on-sapphire and AlN \cite{huang2025acousto,sarabalis2020acousto}, our modulators achieve significantly lower $\mathrm{V_\pi L}$. More notably, even without suspended structures, our devices realize higher $Q_{\mathrm{m}}$ and modulation efficiencies comparable to those of suspended LN platforms \cite{shao2019microwave}, while avoiding their drawbacks such as reduced mechanical stability and limited power-handling capability. In contrast to $\mathrm{As_2S_3}$ based AO modulators \cite{khan2020extraction,wan2022highly,wan2025hybrid}, our platform offers a higher optical Q factor, implying lower optical propagation loss and better compatibility with future integrated photonic components. Furthermore, by enabling efficient excitation of multiple acoustic modes, our modulators can operate across several frequency bands, providing multi-channel AO modulation and paving the way toward wavelength-division multiplexing applications. {While the racetrack resonator demonstrates high efficiency, its footprint (on the order of several hundred micrometers) presents opportunities for further miniaturization to facilitate large-scale integration. To address this, several strategies can be implemented on the LTOI platform. First, utilizing curved interdigital transducers (IDTs) that conform to smaller optical ring resonators (with radii of tens of micrometers) can significantly reduce the acoustic cavity size while maintaining efficient acousto-optic overlap (see Supplementary Note S9). Second, although our work focuses on non-suspended stability, incorporating partial isolation or phononic crystal boundaries could confine acoustic waves in much smaller volumes. Finally, utilizing compact optical Bragg grating cavities instead of racetracks could reduce the device interaction length to a few tens of micrometers.}

Looking forward, future efforts will concentrate on the optimization of optical cavity designs and fabrication processes to further enhance the optical quality factor ($Q_{\mathrm{L}}$). A higher $Q_{\mathrm{L}}$ is expected to significantly improve modulation efficiency by strengthening the photon–phonon interaction within the resonator \cite{wan2025hybrid}. Concurrently, the acoustic resonator will be redesigned to support higher frequency acoustic modes, potentially through heterogeneous integration of LT with other high-acoustic-velocity materials \cite{zhang2023high}. The excitation of higher frequency acoustic waves will enable more efficient optical modulation, paving the way for advanced applications including GHz-FSR optical frequency comb generation, on-chip optical routing, and high-speed optical mode conversion \cite{mantsevich2019optical,li2023frequency,zhang2024integrated}. These developments will further establish LTOI as a versatile platform for high-performance integrated microwave photonic systems.

\section*{Acknowledgments}
The National Key Research and Development Program of China (2022YFA1404601), the National Natural Science Foundation of China (62293520, 62293521, 12074400, 62205363, 12104442, 12404446, 12293052), Shanghai Science and Technology Innovation Action Plan Program (24CL2901000, 24TS1401100, 20JC1416200, 22JC1403300), CAS Project for Young Scientists in Basic Research (Grant No. YSBR-69).\\
The sample fabrication is supported by SIMIT material-device process and characterization platform, ShanghaiTech Material and Device Lab (SMDL).

\section*{Author contributions} 
C.W. and W.Z designed the devices.
X.K. and C.Y. fabricated the LT~wafers.
R.W., B.C. and X.W. fabricated the devices.
C.W. and W.Z analyzed the data.
C.W. and W.Z prepared the figures and wrote the manuscripts with contributions from all authors.
C.W. and X.O. supervised the project.

\section*{Competing interests}
The authors declare no competing financial interests.

\section*{Data Availability Statement} The code and data underlying the results presented in this work are available at Zenodo (https://doi.org/10.5281/zenodo.17131098).

\bibliography{ref_MAIN}	
\bibliographystyle{naturemag}

\end{document}


\title{{\Large Supplementary Information:} \\ High-Efficiency Acousto-Optic Modulation on Non-Suspended Thin-Film Lithium Tantalate}

\author{Weiran Zhou}
\affiliation{School of Microelectronics, University of Science and Technology of China, Hefei, China}
\affiliation{State Key Laboratory of Materials for Integrated Circuits, Shanghai Institute of Microsystem and Information Technology, Chinese Academy of Sciences, 865 Changning Road, Shanghai 200050, China.}

\author{Chengli Wang}
\email[]{wangcl@mail.sim.ac.cn}
\affiliation{State Key Laboratory of Materials for Integrated Circuits, Shanghai Institute of Microsystem and Information Technology, Chinese Academy of Sciences, 865 Changning Road, Shanghai 200050, China.}
\affiliation{Center of Materials Science and Optoelectronics Engineering, University of Chinese Academy of Sciences, Beijing 100049, China}

\author{Xuqiang Wang}
\affiliation{State Key Laboratory of Materials for Integrated Circuits, Shanghai Institute of Microsystem and Information Technology, Chinese Academy of Sciences, 865 Changning Road, Shanghai 200050, China.}
\affiliation{Center of Materials Science and Optoelectronics Engineering, University of Chinese Academy of Sciences, Beijing 100049, China}

\author{Bowen Chen}
\affiliation{State Key Laboratory of Materials for Integrated Circuits, Shanghai Institute of Microsystem and Information Technology, Chinese Academy of Sciences, 865 Changning Road, Shanghai 200050, China.}
\affiliation{Center of Materials Science and Optoelectronics Engineering, University of Chinese Academy of Sciences, Beijing 100049, China}

\author{Jiachen Cai}
\affiliation{State Key Laboratory of Materials for Integrated Circuits, Shanghai Institute of Microsystem and Information Technology, Chinese Academy of Sciences, 865 Changning Road, Shanghai 200050, China.}
\affiliation{Center of Materials Science and Optoelectronics Engineering, University of Chinese Academy of Sciences, Beijing 100049, China}

\author{Tianyao Yang}
\affiliation{State Key Laboratory of Materials for Integrated Circuits, Shanghai Institute of Microsystem and Information Technology, Chinese Academy of Sciences, 865 Changning Road, Shanghai 200050, China.}
\affiliation{Center of Materials Science and Optoelectronics Engineering, University of Chinese Academy of Sciences, Beijing 100049, China}

\author{Dongchen Sui}
\affiliation{State Key Laboratory of Materials for Integrated Circuits, Shanghai Institute of Microsystem and Information Technology, Chinese Academy of Sciences, 865 Changning Road, Shanghai 200050, China.}
\affiliation{Center of Materials Science and Optoelectronics Engineering, University of Chinese Academy of Sciences, Beijing 100049, China}

\author{Xinjian Ke}
\affiliation{Shanghai Novel Si Integration Technology, Shanghai 201815, China}

\author{Yang Chen}
\affiliation{Shanghai Novel Si Integration Technology, Shanghai 201815, China}

\author{Xudong Wang}
\affiliation{State Key Laboratory of Materials for Integrated Circuits, Shanghai Institute of Microsystem and Information Technology, Chinese Academy of Sciences, 865 Changning Road, Shanghai 200050, China.}
\affiliation{Center of Materials Science and Optoelectronics Engineering, University of Chinese Academy of Sciences, Beijing 100049, China}

\author{Ailun Yi}
\affiliation{State Key Laboratory of Materials for Integrated Circuits, Shanghai Institute of Microsystem and Information Technology, Chinese Academy of Sciences, 865 Changning Road, Shanghai 200050, China.}
\affiliation{Center of Materials Science and Optoelectronics Engineering, University of Chinese Academy of Sciences, Beijing 100049, China}

\author{Shibin Zhang}
\affiliation{State Key Laboratory of Materials for Integrated Circuits, Shanghai Institute of Microsystem and Information Technology, Chinese Academy of Sciences, 865 Changning Road, Shanghai 200050, China.}
\affiliation{Center of Materials Science and Optoelectronics Engineering, University of Chinese Academy of Sciences, Beijing 100049, China}

\author{Chengjie Zuo}
\email[]{czuo@ustc.edu.cn}
\affiliation{School of Microelectronics, University of Science and Technology of China, Hefei, China}

\author{Xin Ou}
\email[]{ouxin@mail.sim.ac.cn}
\affiliation{School of Microelectronics, University of Science and Technology of China, Hefei, China}
\affiliation{State Key Laboratory of Materials for Integrated Circuits, Shanghai Institute of Microsystem and Information Technology, Chinese Academy of Sciences, 865 Changning Road, Shanghai 200050, China.}
\affiliation{Center of Materials Science and Optoelectronics Engineering, University of Chinese Academy of Sciences, Beijing 100049, China}
\affiliation{Shanghai Novel Si Integration Technology, Shanghai 201815, China}

\maketitle

\SuppNote{Supplementary Note S1. Calculation of electromechanical coupling coefficient $k^2$ on X-cut lithium tantalate}

The electromechanical coupling coefficient can be expressed as follows \cite{chang2002analysis}:
\begin{equation}
k_{i,j}=\left(\frac{e_{ij}} {\varepsilon_{ii} c_{jj}}\right)^2
\label{eq1}
\end{equation}
where $i=1,2,3...6$ denotes the x, y, z, yz, zx, xy direction in the biaxial crystal. Here, $k_{ij}$ denotes the static electromechanical coupling coefficient when applying an electric field $\mathrm{E}_i$ and a stress field $\mathrm{S}_j$. $e_{ij}, \varepsilon_{ii}$ and $c_{jj}$ are the components of the piezoelectric constant, dielectric constant, and elastic compliance.

The original form and values of tensor elements for $e,\varepsilon~\mathrm{and}~c$ in the crystalline principal axes have been well studied and are given in the references \cite{smith1967elastic} as follows:
\begin{equation*}
e = \begin{bmatrix}
0 & 0 & 0 & 0 & e_{15} & -2e_{22} \\
-e_{22} & e_{22} & 0 & e_{15} & 0 & 0 \\
e_{31} & e_{31} & e_{33} & 0 & 0 & 0 \\
\end{bmatrix},
\end{equation*}
\begin{equation*}
c = \begin{bmatrix}
c_{11} & c_{12} & c_{13} & c_{14} & 0 & 0 \\
c_{12} & c_{11} & c_{13} & c_{14} & 0 & 0 \\
c_{13} & c_{13} & c_{33} & 0 & 0 & 0 \\
c_{14} & c_{14} & 0 & c_{44} & 0 & 0 \\
0 & 0 & 0 & 0 & c_{44} & c_{14} \\
0 & 0 & 0 & 0 & c_{14} & \frac{(c_{11}-c_{12})}{2} \\
\end{bmatrix},
\end{equation*}
\begin{equation}
\varepsilon = \begin{bmatrix}
\varepsilon_{11} & 0 & 0 \\
0 & \varepsilon_{11}& 0 \\
0 & 0 & \varepsilon_{13} \\
\end{bmatrix}.
\label{eq2}
\end{equation}

\begin{figure*}[htbp]
	\centering
	\includegraphics[scale = 1]{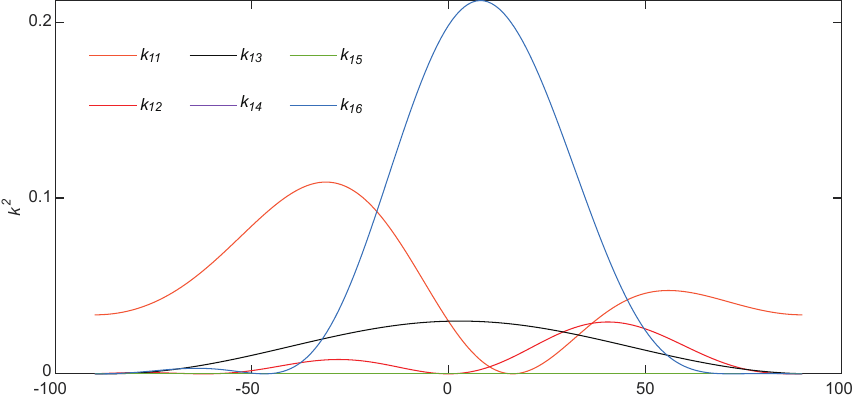}
	\caption{Calculation result of electromechanical coupling coefficient. $k_{11},k_{12},k_{13},k_{14},k_{15},k_{16}$ refer to components of electromechanical coupling coefficient when applying electric field along x axis of device plane.}
	\label{figS1}
\end{figure*}

In the transformed coordinate system $x'$--$y'$--$z'$, obtained by successive rotations of the original frame—first by an angle $\theta$ about the $z$ axis, then by $90^\circ$ about the new $x$ axis, and finally by $90^\circ$ about the resulting $z$ axis—the tensor components are transformed via the transformation matrices $\mathbf{a}$ and $\mathbf{M}$ as follows:
\begin{equation*}
\alpha = \begin{bmatrix}
\cos{\theta}& \sin{\theta} & 0 \\
-\sin{\theta} & \cos{\theta}& 0\\
0 & 0 & 1\\
\end{bmatrix}\times
\begin{bmatrix}
1 & 0 & 0 \\
0 & 0& 1 \\
0& -1 & 0 \\
\end{bmatrix}\times
\begin{bmatrix}
0 & 1 & 0 \\
-1 & 0& 0 \\
0 & 0 & 1 \\
\end{bmatrix}=
\begin{bmatrix}
0& \cos{\theta} & \sin{\theta} \\
0&-\sin{\theta} & \cos{\theta}\\
1 & 0 & 0\\
\end{bmatrix}
\end{equation*}
\begin{equation*}
M = \begin{bmatrix}
0 & \cos^{2}\theta & \sin^{2}\theta & 2\cos\theta\sin\theta & 0 & 0 \\
0 & \sin^{2}\theta & \cos^{2}\theta & -2\sin\theta\cos\theta & 0 & 0 \\
1 & 0 & 0 & 0 & 0 & 0 \\
0 & 0 & 0 & 0 & \cos\theta & -\sin\theta \\
0 & 0 & 0 & 0 & \sin\theta & \cos\theta \\
0 & -\cos\theta\sin\theta & \cos\theta\sin\theta & \cos^{2}\theta - \sin^{2}\theta & 0 & 0
\end{bmatrix}
\end{equation*}
\begin{equation}
e'=aeM^T;\quad c'=McM^T;\quad \varepsilon'=a\varepsilon a^T.
\label{eq3}
\end{equation}

Substituting Eq.~\ref{eq2} and Eq.~\ref{eq3} into Eq.~\ref{eq1} yields the electromechanical coupling coefficient of lithium tantalate, as illustrated in Fig.~\ref{figS1}. The resulting curve indicates that the excitation of surface acoustic waves in lithium tantalate is predominantly influenced by $k_{11}$ and $k_{16}$, which correspond to the electromechanical coupling coefficients of the Rayleigh and shear modes, respectively.
\\
 {To contextualize the performance of the LTOI platform, we compare its properties with other piezoelectric materials commonly used in integrated photonics. While Lithium Niobate (LN) exhibits the highest electromechanical coupling coefficient ($k^2$), Lithium Tantalate (LT) distinguishes itself with superior acoustic Quality Factors ($Q$) and better thermal stability, characterized by a lower Temperature Coefficient of Frequency (TCF) \cite{takai2017ihp,tanaka2024evolution}. As summarized in Table \ref{tab:material_comparison}, although LT's $k^2$ is slightly lower than that of LN, it offers a balanced combination of strong piezoelectric coupling (significantly higher than AlN and GaAs) and excellent acoustic coherence. This intrinsic material advantage makes LTOI particularly suitable for acousto-optic devices where thermal stability and low acoustic loss are prioritized alongside modulation efficiency.}

\begin{table}[htbp]
    \centering
    \caption{Comparison of piezoelectric properties, acoustic quality, and thermal stability of common integrated photonic materials.}
    \label{tab:material_comparison}
    \begin{tabular}{lcccc}
        \toprule
        \textbf{Material} & \textbf{Effective $k^2$ (\%)} & \textbf{Acoustic $Q$} & \textbf{TCF (ppm/$^\circ$C)} & \textbf{Primary Application} \\
        \midrule
        \textbf{Lithium Tantalate (LT)} & \textbf{5 -- 21} & \textbf{High} & \textbf{$-30$ to $-45$} & \textbf{Stable RF Filters, AO} \\
        Lithium Niobate (LN) & 5 -- 40 & Moderate & $-70$ to $-90$ & Wideband Filters, Modulators \\
        Aluminum Nitride (AlN) & 0.3 -- 1.0 & High & $-25$ & MEMS, FBAR Filters \\
        Gallium Arsenide (GaAs) & $\sim$ 0.07 & Low/Mod & $-70$ to $-90$ & Active Photonics \\
        \bottomrule
    \end{tabular}
    \vspace{2mm}
    
    \footnotesize{Note: Values represent typical ranges for surface acoustic wave (SAW) modes. LT demonstrates a superior Temperature Coefficient of Frequency (TCF) compared to LN \cite{Bartasyte2012}. AlN and GaAs values are referenced from \cite{Piazza2006} and \cite{Adachi1985} respectively.}
\end{table}

\SuppNote{Supplementary Note S2. FEM simulation of SAW with different propagation angle}

\begin{figure*}[htbp]
	\centering
	\includegraphics[scale = 1]{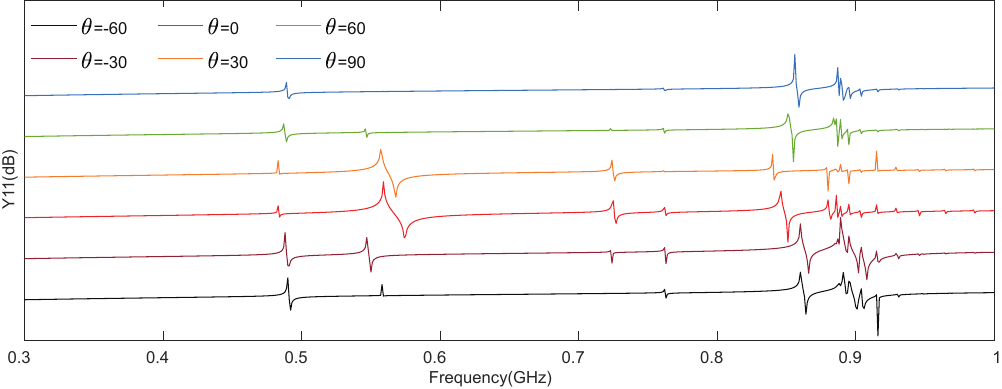}
	\caption{FEM simulation of admittance curve of acoustic resonator with different propagation angle on LTOI.}
	\label{figS2}
\end{figure*}

 {Specifically, the 3D FEM simulations were conducted using a unit cell representing a single period of the IDT electrode structure to match the target wavelength. Periodic boundary conditions (PBC) were applied along the propagation direction to simulate an infinite periodic structure. This method avoids the massive computational resources required to mesh the entire resonant cavity (comprising hundreds of IDT finger pairs) while enabling the precise extraction of phase velocities and cross-sectional mode shapes used for calculating the acousto-optic overlap integral. Utilizing this model, FEM simulations were performed to investigate the influence of propagation angle on the excitation efficiency of surface acoustic wave (SAW) modes.} The results are presented in \fref{figS2}. As evidenced by the obtained curves, the R0 mode is most efficiently excited at a propagation angle of $\theta = -30^\circ$. {As evidenced by the obtained curves, the R0 mode is most efficiently excited at a propagation angle of $\theta=-30^{\circ}$. Conversely, a distinct suppression of the R0 mode is observed at $\theta=30^{\circ}$. This angular selectivity is mathematically attributed to the transformation of the piezoelectric tensor components in the rotated coordinate system, which results in a vanishing effective piezoelectric coefficient specifically at this orientation.} In contrast, the SH0 mode exhibits optimal excitation efficiency at $\theta = 0^\circ$. The R1 mode demonstrates high excitation efficiency across a range of angles, with the notable exception of $\theta = 30^\circ$.

\SuppNote{Supplementary Note S3. Illustration of the fabrication processes and characterization of AOM}

\begin{figure*}[htbp]
	\centering
	\includegraphics[scale = 1]{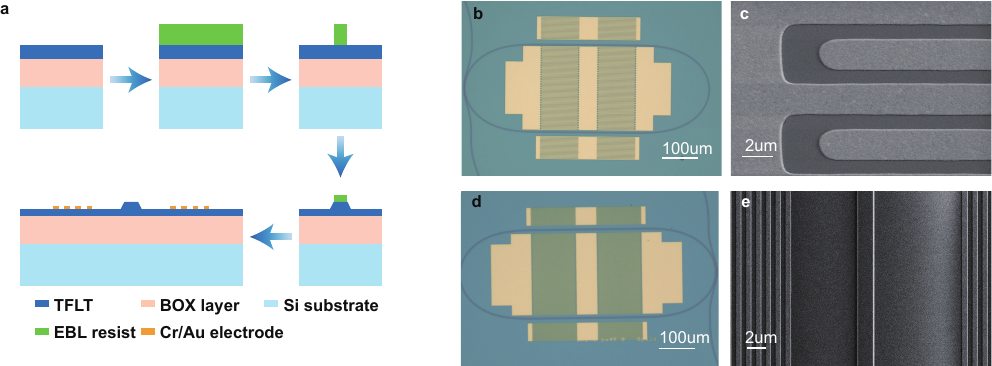}
	\caption{\textbf{Fabrication and characterization of AOM.} (a) Schematic of the fabrication process of AOM. (b) Partial microscope image of the RT AOM at low frequency. (c) Partial SEM image of RT AOM at low frequency. (d) Partial microscope image of the RT AOM at high frequency. (e) Partial SEM image of RT AOM at high frequency.}
	\label{figS3}
\end{figure*}

 {Before device fabrication, the LTOI stack was prepared at the SIMIT material-device process platform. An optical grade X-cut $\mathrm{LiTaO}_3$ wafer was implanted with He ions... [briefly describe parameters]. The wafer was then bonded to a $\mathrm{SiO}_2/\mathrm{Si}$ substrate... After splitting, the thin film was annealed and polished to remove the damage layer, resulting in a \SI{600}{\nano\meter} single-crystal thin film.}

The fabrication process for the acousto-optic modulators (AOMs) is illustrated in \fref{figS3}(a). Using this process, a racetrack (RT) AOM with an acoustic wavelength of \SI{6}{\micro\meter} was fabricated. Figure \ref{figS3}(b) shows an optical micrograph of the acousto-optic modulator, while \fref{figS3}(c) provides a detailed view of the IDT electrodes. {Specifically, the IDT consists of 30 finger pairs with a uniform period and finger width throughout. No apodization or chirping was applied to the inner or outer electrodes. This uniformity ensures that the excited acoustic wavefront maintains a consistent phase relation, allowing the high-Q Fabry-Perot cavity formed by the external reflectors to effectively confine the acoustic energy and dominate the resonant response.} For the microwave photonic link application discussed in the main text, an additional RT AOM with a reduced acoustic wavelength of \SI{2}{\micro\meter} was fabricated following the same process flow. To achieve the finer IDT electrode features required at this scale, the maskless laser direct writing (MLA) alignment process described above was replaced with electron-beam lithography (EBL) alignment. A higher-resolution electron-beam resist, ZEP520A, was employed to accurately pattern the nanoscale electrode structures. \fref{figS3}(d) presents an overall optical micrograph of this device, and \fref{figS3}(e) shows a scanning electron microscopy (SEM) image detailing the IDT and optical waveguide structures.

\SuppNote{Supplementary Note S4. Optimization of AOM}

\begin{figure*}[htbp]
	\centering
	\includegraphics[scale = 1]{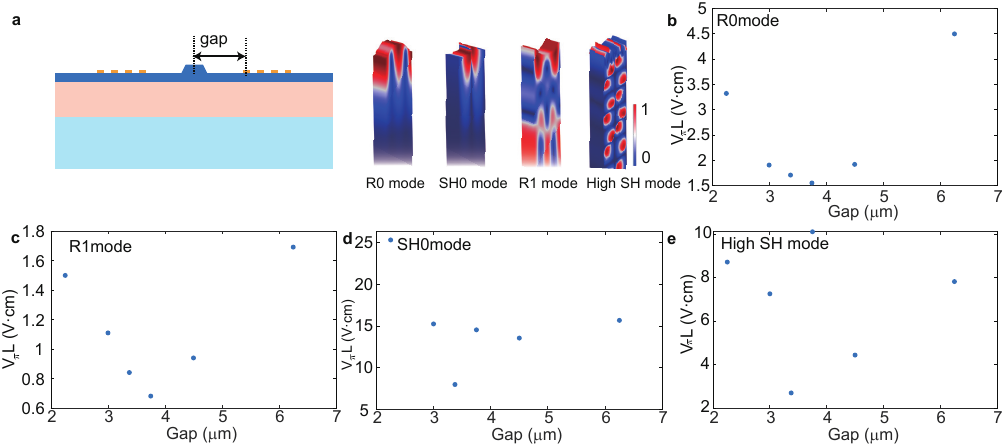}
	\caption{\textbf{Optimization of waveguide-electrode spacing.} (a) Experimental setup for AOMZI characterization. (b) Modulation efficiency of R0 mode versus spacing. (c) Modulation efficiency of R1 mode versus spacing. (d) Modulation efficiency of SH0 mode versus spacing. (e) Modulation efficiency of higher-order SH modes versus spacing.}
	\label{figS4}
\end{figure*}

To further optimize the performance of the acousto-optic modulator, the spacing between the electrode and the waveguide {was systematically varied to maximize the acousto-optic coupling efficiency. This optimization is essential to achieve the push-pull condition described in the main text. By adjusting the gap, we ensure that the acoustic phase difference between the two waveguide arms approximates $\pi$ ($180^\circ$), thereby maximizing the differential phase shift.} A series of Mach–Zehnder interferometer-type acousto-optic modulators (AOMZIs) with acoustic wave propagation along the Z-axis of lithium tantalate were fabricated and characterized using the same experimental setup as described in the main text \fref{figS4}(a). The modulation efficiency of each AOMZI was measured across the frequency ranges corresponding to different acoustic modes while varying the electrode-waveguide spacing. The R1 mode, which exhibited the strongest acousto-optic coupling, achieved optimal modulation efficiency at a spacing of \SI{3.375}{\micro\meter}, as summarized in \fref{figS4}(c). A similar trend was observed for both the R0 and SH0 modes \fref{figS4}(b,d), which can be attributed to their nearly identical acoustic wavelengths and mode profiles along the X-axis of the device. To further examine this dependence, higher-order shear-horizontal (SH) modes were also characterized \fref{figS4}(e), revealing a similar behavior due to their similar mode extent and energy distribution in the X axis. These results demonstrate that modulation efficiency can be substantially enhanced through precise optimization of the electrode-waveguide spacing.

\SuppNote{Supplementary Note S5. Anisotropy of R1 mode}

\begin{figure*}[ht]
	\centering
	\includegraphics[scale = 1]{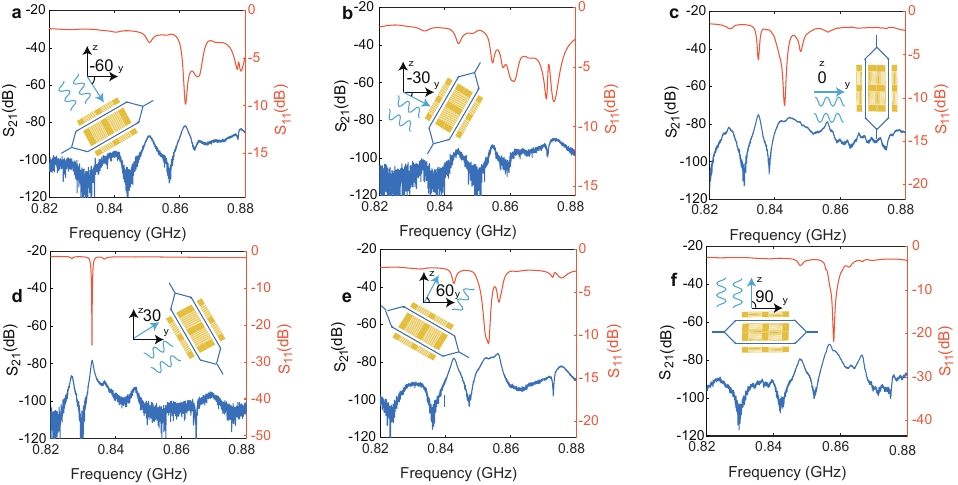}
	\caption{\textbf{Characterization of the anisotropy of AO modulation in Lithium Tantalate.}
 (a) Measured $\mathrm{S_{11}}$ and $\mathrm{S_{21}}$ spectrum of R0 mode when $\theta$~=~-60$\degree$, the optical power detected by the photoreceiver is 0.16 mW. (b) Measured $\mathrm{S_{11}}$ and $\mathrm{S_{21}}$ spectrum of R0 mode when $\theta$~=~-30$\degree$, the optical power detected by the photoreceiver is 0.12mW. (c) Measured $\mathrm{S_{11}}$ and $\mathrm{S_{21}}$ spectrum of R0 mode when $\theta$~=~0$\degree$, the optical power detected by the photoreceiver is 0.2 mW. (d) Measured $\mathrm{S_{11}}$ and $\mathrm{S_{21}}$ spectrum when $\theta$~=~30$\degree$, The optical power detected by the photoreceiver is 0.1 mW. (e) Measured $\mathrm{S_{11}}$ and $\mathrm{S_{21}}$ spectrum when $\theta$~=~60$\degree$, The optical power detected by the photoreceiver is 0.11 mW. (f) Measured $\mathrm{S_{11}}$ and $\mathrm{S_{21}}$ spectrum when $\theta$~=~90$\degree$, The optical power detected by the photoreceiver is 0.1 mW.}
	\label{figS5_aniso}
\end{figure*}

Based on the standard extraction method of
$\mathrm{V_\pi L}$~for MZI AO modulators, the values of R0 mode for devices oriented at –60 $\degree$, –30$\degree$, 0$\degree$, 30$\degree$, 60$\degree$ and 90$\degree$ were determined to be 2.649, 5.954, 1.7865, 1.358, 1.0385, and 0.68 $\mathrm{V \cdot cm}$.

 {Note that unlike the R0 mode, the effective $k^2$ values for the R1 mode presented in Fig. 3(h) of the main text are derived from FEM simulations. As discussed in the main text, the experimental extraction of $k^2$ for R1 was precluded by the presence of dense longitudinal modes within the high-Q acoustic cavity, which obscured the distinct series and parallel resonance frequencies.}

\SuppNote{Supplementary Note S6. Characterization of optical cavity}

\begin{figure*}[htbp]
	\centering
	\includegraphics[scale = 1]{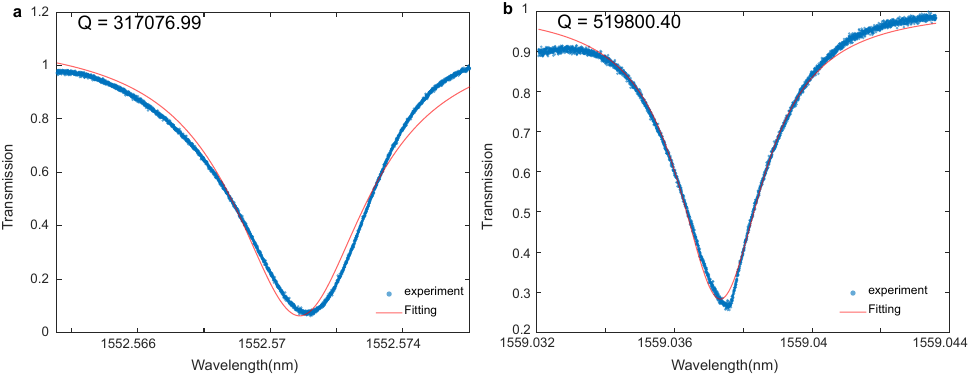}
	\caption{\textbf{Measurement of the racetrack cavity.}~(a) Transmission of racetrack cavity used in the low frequency AOM. (b) Transmission of racetrack cavity used in the high frequency AOM.}
	\label{figS6}
\end{figure*}

Based on the fabricated optical racetrack resonators, wavelength scanning was performed to obtain the transmission spectra. Lorentzian fitting was applied to the measured resonance dips, as presented in \fref{figS6}(a) and \fref{figS6}(b) for the racetrack cavities corresponding to \fref{figS3}(b) and \fref{figS3}(d), respectively. The extracted loaded quality factors ($Q_L$) were $3.17 \times 10^5$ and $5.20 \times 10^5$.

 {During the AO modulation measurements (Main text Figs. 4 and 5), the recorded optical powers detected by the PD were 0.13 mW (low frequency device) and 0.08 mW (high frequency device), respectively.}

\SuppNote{Supplementary Note S7. Analysis of Acoustic Cavity, Degenerated Modes, and Angle Dependence}

\begin{figure*}[htbp]
	\centering
	\includegraphics[scale = 1]{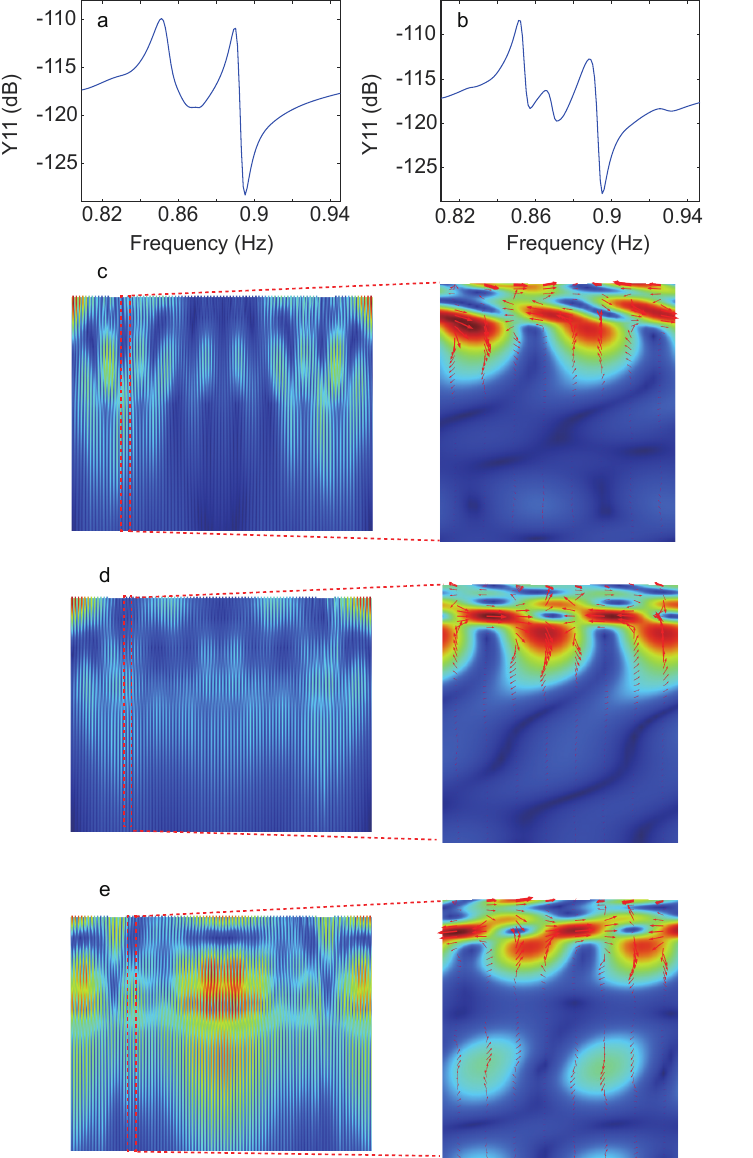}
	\caption{\textbf{FEM simulation of the acoustic cavity effect.} The model incorporates 30 pairs of IDT electrodes and 10 pairs of reflector grating electrodes. (a) Simulated admittance ($Y_{11}$) curve \textit{without} acoustic reflectors, showing two intrinsic modes. (b) Simulated $Y_{11}$ curve \textit{with} acoustic reflectors, showing the appearance of a new cavity mode between the intrinsic ones. (c) Mode profile of the main $R_1$ resonance at 0.852 GHz (Intrinsic). (d) Mode profile of the reflector-induced cavity mode at 0.865 GHz (New). (e) Mode profile of the intrinsic spatially degenerate mode at 0.889 GHz (Intrinsic).}
	\label{figS7}
\end{figure*}

To elucidate the physical origin of the fine spectral structures ("degenerated modes") observed in the RF measurements and their dependence on propagation angle, we performed finite element method (FEM) simulations based on the fabricated device geometry. The simulation model incorporates \textbf{30 pairs of IDT electrodes and 10 pairs of reflector grating electrodes}.

\textbf{1. Origin of Spectral Splitting:}
As shown in \fref{figS7}(a), the simulated admittance ($Y_{11}$) curve for the device \textit{without} the grating structure already exhibits two distinct resonance peaks. The corresponding displacement profiles indicate that:
\begin{itemize}
  \item The peak at 0.852 GHz corresponds to the main $R_1$ mode (\fref{figS7}(c)).
  \item The peak at 0.889 GHz corresponds to a secondary intrinsic mode arising from spatial variations in the vibration distribution (\fref{figS7}(e)).
\end{itemize}

Upon introducing the grating electrodes, as shown in \fref{figS7}(b), the spectral response evolves. A new resonance peak emerges at 0.865 GHz, located exactly between the two intrinsic modes. The mode profile of this specific peak, presented in \fref{figS7}(d), confirms that it is a \textbf{longitudinal cavity mode} confined effectively by the acoustic reflectors. Thus, the experimentally observed splitting is a superposition of intrinsic spatial degeneracy and the enhanced cavity mode.

\begin{figure*}[htbp]
	\centering
	\includegraphics[scale = 1]{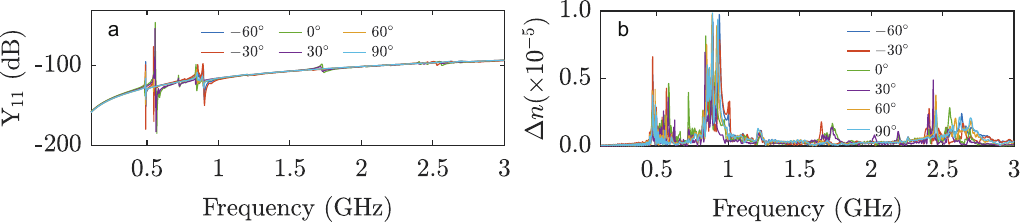}
	\caption{\textbf{Quantitative analysis of acousto-optic coupling.} (a) Simulated $Y_{11}$ admittance curves of the whole AOM structure with different in-plane propagation angles. (b) Calculated refractive index change ($\Delta n$) as a function of frequency, based on a microwave input power of 0.001 W. The peaks in (b) correspond to the acoustic resonant modes identified in (a). The simulation results indicate that at a propagation angle of $\theta=90^\circ$, the refractive index modulation reaches its maximum at 0.887 GHz, corresponding to the R1 mode.}
	\label{figS8}
\end{figure*}

\textbf{2. Angle Dependence of Visibility:}
The experimental observation that these extra resonances are prominent only at specific angles (as noted in Fig. 3) is attributed to the strong anisotropy of X-cut LiTaO$_3$.
The high-Q cavity mode (\fref{figS8}(a)) requires efficient acoustic reflection and strong electromechanical transduction to be observed. At optimal propagation angles (e.g., near $\theta = -30^\circ$ for the Rayleigh mode), the effective electromechanical coupling coefficient ($k_{eff}^2$) is maximized, allowing this discrete longitudinal mode to be strongly excited and spectrally resolved. Conversely, at off-axis angles where coupling is weak or leakage is high, the cavity mode is not efficiently driven, causing the fine spectral splitting to broaden or disappear from the measured spectrum.

\SuppNote{Supplementary Note S8. Quantitative analysis of acousto-optic overlap integral}

We appreciate the suggestion to quantify the mode overlap. To address this, we calculated the acousto-optic overlap integral as a format of change of refractive index for the different acoustic modes (R0, R1, SH0) with the fundamental TE optical mode. The overlap integral is defined as:

\begin{equation}
  \Delta n = \Delta n_{pe} + \Delta n_{eo} + \Delta n_{mb}
\end{equation}

Where $\Delta n_{pe}$, $\Delta n_{eo}$, $\Delta n_{mb}$ refer to refractive index change caused by photo-elastic effect, electro-optic effect (electric field comes from acoustic wave through piezoelectric effect) and moving boundary effect. The calculation of these effects can be done by:

\begin{equation}
  \Delta n_{pe,eo} = \frac{\epsilon_0 n^5}{2} \frac{\int dr (E_x^*, E_y^*, E_z^*) \begin{pmatrix} dB_1 & dB_6 & dB_5 \\ dB_6 & dB_2 & dB_4 \\ dB_5 & dB_4 & dB_3 \end{pmatrix} \begin{pmatrix} E_x \\ E_y \\ E_z \end{pmatrix}}{\int \vec{E}^* \epsilon \vec{E} dr}
\end{equation}

\begin{equation}
  \Delta n_{mb} = -\frac{n}{2} \frac{\oint (\vec{R} \cdot \vec{n}) \left( \Delta \epsilon (\vec{E}^* \times \vec{n} \cdot \vec{E} \times \vec{n}) - \Delta \epsilon^{-1} (\vec{D}^* \cdot \vec{n})(\vec{D} \cdot \vec{n}) \right) dS}{\int \vec{E}^* \epsilon \vec{E} dr}
\end{equation}

Where $\vec{E}$ denotes that electric field of optic wave. $R$ refers to displacement field. $\vec{D}$ is the electric displacement field of optical mode. To further numerically calculate the index change, the coefficient can be written as below:

\begin{equation}
  \begin{pmatrix} dB_{1_{pe}} \\ dB_{2_{pe}} \\ dB_{3_{pe}} \\ dB_{4_{pe}} \\ dB_{5_{pe}} \\ dB_{6_{pe}} \end{pmatrix} =
  \begin{pmatrix}
  p_{33} & p_{31} & p_{31} & 0 & 0 & 0 \\
  p_{13} & p_{11} & p_{12} & 0 & p_{14} & 0 \\
  p_{13} & p_{12} & p_{11} & 0 & -p_{14} & 0 \\
  0 & 0 & 0 & p_{66} & 0 & p_{14} \\
  0 & p_{41} & -p_{41} & 0 & p_{44} & 0 \\
  0 & 0 & 0 & p_{41} & 0 & p_{66}
  \end{pmatrix}
  \begin{pmatrix} S_1 \\ S_2 \\ S_3 \\ S_4 \\ S_5 \\ S_6 \end{pmatrix}
\end{equation}

\begin{equation}
  \begin{pmatrix} dB_{1_{eo}} \\ dB_{2_{eo}} \\ dB_{3_{eo}} \\ dB_{4_{eo}} \\ dB_{5_{eo}} \\ dB_{6_{eo}} \end{pmatrix} =
  \begin{pmatrix}
  r_{33} & 0 & 0 \\
  r_{13} & 0 & -r_{22} \\
  r_{13} & 0 & -r_{22} \\
  0 & -r_{22} & 0 \\
  0 & 0 & r_{51} \\
  0 & -r_{51} & 0
  \end{pmatrix}
  \begin{pmatrix} \varepsilon_x \\ \varepsilon_y \\ \varepsilon_z \end{pmatrix}
\end{equation}

where $p_{ij}$, $r_{ij}$ denote photoelastic and electro-optic coefficient. And $\vec{\varepsilon}$ refers to electric field induced by piezoelectric effect from acoustic wave.

For the quantitative calculation of the refractive index change, the simulation was performed with a microwave input power of 0.001 W. The resonant frequencies for these calculations were identified based on the admittance curves ($Y_{11}$) shown in \textbf{Fig. S8(a)}.

Our calculations, as illustrated in Fig. S8, reveal that at a propagation angle of $\theta=90^\circ$, the maximum refractive index change occurs at a frequency of 0.887 GHz, which corresponds to the R1 mode. This overlap factor is significantly higher than that of the R0 and SH0 modes. This quantitative analysis confirms that the R1 mode profile has the highest spatial overlap with the optical mode within the waveguide core, leading to the superior modulation efficiency observed experimentally.

\SuppNote{Supplementary Note S9. Theoretical analysis of modulation sidebands in the optical spectrum}

To further elucidate the origin of the multiple resonance dips observed in the optical transmission spectrum (Fig. 5d of the main text), we provide a theoretical description of the acousto-optic (AO) modulation process. When the AO modulator is driven by a strong radio-frequency (RF) signal, the optical carrier at frequency $f_{opt}$ is phase-modulated by the acoustic field at frequency $f_a$. The modulated electric field $E(t)$ can be expressed using the Jacobi-Anger expansion:

\begin{equation}
E(t) = E_0 e^{i\omega_0 t} \sum_{n=-\infty}^{\infty} J_n(\beta) e^{in\Omega t}
\end{equation}

where $\omega_0 = 2\pi f_{opt}$ is the optical carrier angular frequency, $\Omega = 2\pi f_a$ is the acoustic angular frequency, $J_n$ denotes the $n$-th order Bessel function of the first kind, and $\beta$ represents the modulation depth. The modulation depth $\beta$ is proportional to the square root of the applied RF power.
In a high-$Q$ optical resonator ($Q_L \approx 5.2 \times 10^5$), these generated sidebands ($f_{opt} \pm n \cdot f_a$) align with the cavity resonance modes during the laser wavelength scanning process. As the RF power increases, the energy transfer from the optical carrier to these sidebands becomes more efficient, as determined by the power distribution $P_n \propto |J_n(\beta)|^2$. Consequently, the sideband dips in the transmission spectrum become deeper and more numerous. This behavior confirms the high modulation efficiency and strong photon-phonon interaction achieved on the non-suspended LTOI platform without requiring additional optical amplification.

\SuppNote{Supplementary Note S10. Simulation of Ring-shaped Acoustic Resonator}

\begin{figure*}[htbp]
	\centering
	\includegraphics[scale = 0.8]{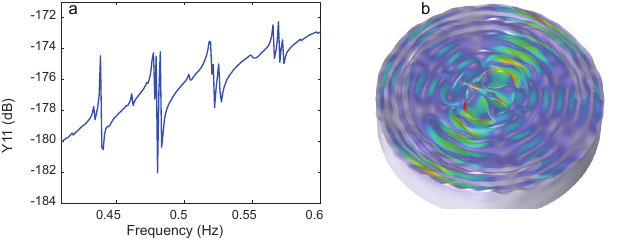}
	\caption{\textbf{Simulation of ring-shaped acoustic resonator.} (a) Calculated admittance curve ($Y_{11}$) obtained from FEM simulation of the ring-shaped acoustic resonator. (b) Displacement field distribution of the acoustic mode at a frequency of 0.438 GHz.}
	\label{figS9}
\end{figure*}

To demonstrate the feasibility of device miniaturization, we performed finite element method (FEM) simulations of a ring-shaped acoustic resonator. As illustrated in \fref{figS9}, ring-shaped interdigital transducers (IDTs) were designed to excite acoustic waves that conform to a circular optical micro-ring. The simulated admittance curve, shown in \fref{figS9}(a), indicates the effective excitation of acoustic modes. The displacement field distribution at the resonance frequency of \SI{0.438}{\giga\hertz} is presented in \fref{figS9}(b). The results confirm that the acoustic energy can be sufficiently confined within the ring structure, suggesting that the device footprint can be significantly reduced compared to the racetrack topology while maintaining sufficient overlap for modulation.

\SuppNote{Supplementary Note S11. Theoretical analysis of detuning dependence}

To understand the dependence of acousto-optic modulation efficiency on cavity detuning, we consider the theory of cavity optomechanics. The optomechanical interaction induces an additional optical damping rate ($\Gamma_{opt}$) on the mechanical mode, which modifies the intrinsic mechanical damping rate. This optical damping rate is a function of the detuning $\Delta = \omega_{laser} - \omega_{cavity}$ and is given by \cite{jiang2019lithium}:

\begin{equation}
\Gamma_{opt}(\Delta) = g_0^2 N \left( \frac{\kappa}{(\Delta - \Omega_m)^2 + (\kappa/2)^2} - \frac{\kappa}{(\Delta + \Omega_m)^2 + (\kappa/2)^2} \right)
\end{equation}

where $g_0$ is the vacuum optomechanical coupling rate, $N$ is the intracavity photon number, $\kappa$ is the cavity decay rate, and $\Omega_m$ is the mechanical frequency.

\begin{itemize}
  \item \textbf{Red Detuning ($\Delta < 0$):} The first term dominates, leading to $\Gamma_{opt} > 0$. This results in additional damping (cooling) of the mechanical motion, which suppresses the signal in modulation applications.
  \item \textbf{Blue Detuning ($\Delta > 0$):} The second term dominates, leading to $\Gamma_{opt} < 0$. This negative damping indicates a gain mechanism where the mechanical oscillation is amplified (phonon creation/lasing), thereby enhancing the modulation efficiency observed in the $S_{21}$ measurements.
\end{itemize}

\bibliography{ref_SI}
\bibliographystyle{naturemag}